\documentclass[aps,12pt,a4paper]{revtex4}
\usepackage{epsfig}
\usepackage{graphicx}
\usepackage{amsmath,amssymb,color}

\parskip=\medskipamount

\newcommand{\eq}[1]{(\ref{#1})}
\newcommand{\fig}[1]{Fig.\ref{#1}}

\newcommand{\be}{\begin{equation}}
\newcommand{\ee}{\end{equation}}

\newcommand\disp{\displaystyle}

\newcommand{\eps}{\varepsilon}

\begin{document}

\title{Two conjectures about spectral density of diluted sparse Bernoulli random matrices}

\author{S.K. Nechaev$^{1,3,5}$}

\address{$^{1}$Universit\'e Paris-Sud/CNRS, LPTMS, UMR8626, B\^at. 100, 91405 Orsay, France, \\
$^{3}$Department of Applied Mathematics, National Research University Higher School of Economics,
101000, Moscow, Russia, \\ $^{5}$P.N. Lebedev Physical Institute of the Russian Academy of
Sciences, 119991, Moscow, Russia.}

\begin{abstract}

We consider the ensemble of $N\times N$ ($N\gg 1$) symmetric random matrices with the bimodal
independent distribution of matrix elements: each element could be either "1" with the probability
$p$, or "0" otherwise. We pay attention to the "diluted" sparse regime, taking $p=1/N +\eps$, where
$0<\eps \ll 1/N$. In this limit the eigenvalue density, $\rho(\lambda)$, is essentially singular,
consisting of a hierarchical ultrametric set of peaks. We provide two conjectures concerning the
structure of $\rho(\lambda)$: (i) we propose an equation for the position of sequential (in
heights) peaks, and (ii) we give an expression for the shape of an outbound enveloping curve. We
point out some similarities of $\rho(\lambda)$ with the shapes constructed on the basis of the
Dedekind modular $\eta$-function.

\end{abstract}

\maketitle

\section{Preliminaries and conjectures}

The Bernoulli matrix model considered in this letter, is defined as follows. Take a large $N\times
N$ symmetric matrix $A$ ($N\gg 1$) with the matrix elements, $a_{ij}=a_{ji}$, independent
identically distributed random variables, equal to "1" with probability $p$ for any $i\neq j$, and
to "0" otherwise. This defines the uniform distribution on the entries $a_{ij}$:
\be
{\rm Prob}(a_{ij})=p \delta(a_{ij}-1) + (1-p) \delta(a_{ij}),
\label{eq:01}
\ee
where $\delta$ is the Kronecker symbol: $\delta(x)=1$ for $x=0$, and $\delta(x)=0$ otherwise. The
matrix $A$ can be regarded as an adjacency matrix of a random Erd\"os-R\'enyi graph without
self-connections. Obviously, all the eigenvalues, $\lambda_n$ ($n=1,...N$) of the matrix $A$ are
real.

\begin{figure}[ht]
\epsfig{file=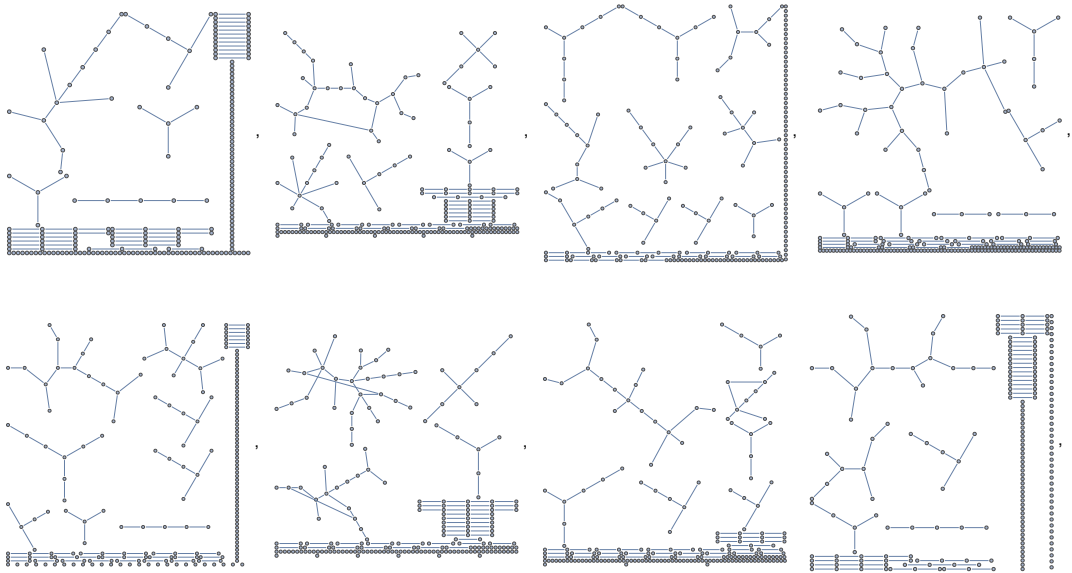,width=14cm}
\caption{Few typical samples of collections of graphs of randomly generated $200\times 200$ adjacency
Bernoulli matrix $A$ at $p=0.00502 = 1/N+2\times 10^{-5}$.}
\label{fig:1}
\end{figure}

Let $\rho(\lambda)$ be the eigenvalue density in the ensemble of $A$. For $N\gg 1$ the limiting
shape of $rho(\lambda)$ is known in various cases. If $p$ is large enough (of order of unity), then
for $N\gg 1$ the function $\rho(\lambda)$ tends to the Wigner semicircle law,
$\sqrt{4N-\lambda^2}$, typical for the Gaussian matrix ensembles. For $p=c/N$ ($c>1$), the matrix
$A$ is \emph{sparse} and the density $\rho(\lambda)$ in the ensemble of sparse matrices has
singularities at finite values of $c$ \cite{rod1,rod2,fyod}. At $c=1$ one has in average of order
of one nonzero element in any row of $A$ and therefore below $c=1$ the entire matrix becomes a
collection of almost disjoint elements. This results in the trivial spectral density,
$\rho(\lambda)=\delta(\lambda)$, in the ensemble of $A$. In the works \cite{evan,bauer,sem,kuch}
the behavior of the spectral density, $\rho(\lambda)$, has been analyzed in the limit when $c$
tends to unity. It has been pointed out that the function $\rho(\lambda)$ becomes more and more
singular as $c$ approaches 1. Slightly above 1, the typical subgraphs of the random matrix $A$, are
basically random linear chains or disjoint trees and the eigenvalues of the entire matrix $A$ are
given by corresponding characteristic polynomials of these simple graphs. Few typical samples of
randomly generated by \emph{Mathematica} $200\times 200$ matrices $A$ at $p=0.00502 = 1/N+2\times
10^{-5}$ (i.e. for $c=1.004$) are shown in the \fig{fig:1}. Note the essential fraction of linear
subgraphs. The trees with loops appear rarely.

The eigenvalues contributing to the spectrum of $A$, for example, from a 3-star graph, are obtained
from the following equation
\be
\det\left(
\begin{array}{cccc}
-\lambda & 1 & 0 & 0 \\  1 & -\lambda & 1 & 1 \\  0 & 1 & -\lambda & 0 \\  0 & 1 & 0 & -\lambda
\end{array} \right) = 0
\label{eq:02}
\ee
whose solution is $\lambda=\{0,0,\sqrt{3},\sqrt{3}\}$.

\begin{figure}[ht]
\epsfig{file=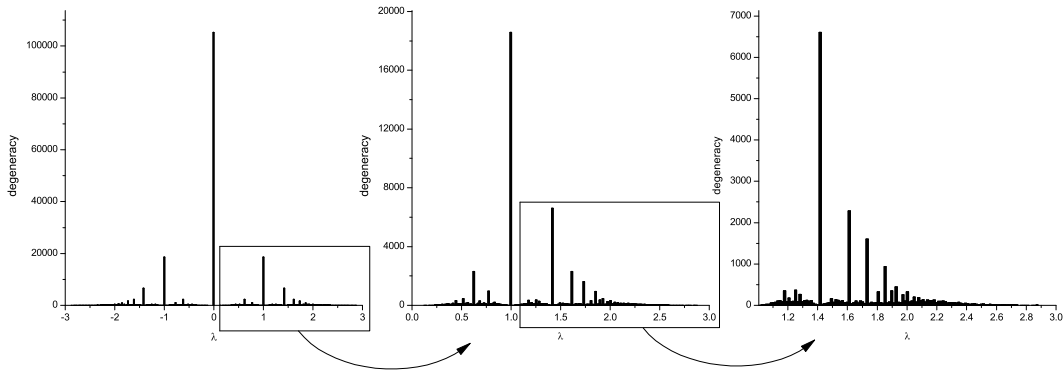,width=14cm}
\caption{Ultrametric hierarchical structure of a spectral density, $\rho(\lambda)$, for ensemble
of 1000 symmetric random $200\times 200$ Bernoulli matrices, generated at $p=0.00502 = 1/N+2\times
10^{-5}$.}
\label{fig:2}
\end{figure}

The whole spectrum of a symmetric matrix $A$ consists of singular peaks, located at $\lambda_i$
($i=1,..N$), while the heights of peaks are the multiplicities of $\lambda_i$. The singular
spectral density, $\rho(\lambda)$, for an ensemble of 1000 random symmetric matrices $A$, each of
size $200\times 200$ and generated with the probability $p=0.00502$, is shown in the \fig{fig:2}.
One can note that the spectrum possess the ultrametric hierarchical structure. In this letter we
conjecture some statistical properties of the spectral density, $\rho(\lambda)$, of ensemble of
sparse random symmetric matrices in the limit $p=1/N+\eps$, where $0<\eps\ll 1/N$.

\begin{figure}[ht]
\epsfig{file=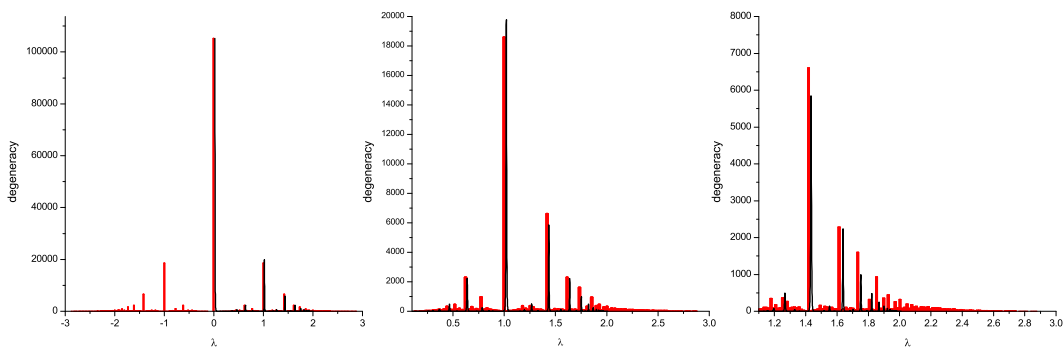,width=14cm}
\caption{Numerical data, shown in red is as in the \fig{fig:2} and the analytic curve, shown in
black is given by \eq{eq:comp1}--\eq{eq:comp2}. The small shift in $\Delta \lambda = +0.02$ between
red and black curves is specially inserted by hands to make data distinguishable.}
\label{fig:2-1}
\end{figure}

The singular spectral density depicted in the \fig{fig:2} is compared in the \fig{fig:2-1} with the
function $g(\lambda)$, defined as follows
\be
g(\lambda) = \left[-\ln f\left(\frac{1}{\pi}\arccos \frac{\lambda}{2}\Big| y\right)\right]^2 \qquad
(0<y\ll 1)
\label{eq:comp1}
\ee
where
\be
f(x|y) = A |\eta(x+i y)| \qquad (-2<x<2,\; y>0)
\label{eq:comp2}
\ee
is the Dedekind $\eta$-function ($A$ is some normalization constant). Since the spectral density
$\rho(\lambda)$ is symmetric, later on we inspect only the part $\lambda\ge 0$.

As one sees from the \fig{fig:2-1} the correspondence between our numerical eigenvalue counting for
random symmetric sparse Bernoulli matrices $A$ with the function $g(x)$, defined in
\eq{eq:comp1}--\eq{eq:comp2}, is good in the bulk (we have specially introduced the small shift in
$\Delta \lambda = 0.02$ between numerical data and analytic expression to make them
distinguishable). One also sees that the tails of the spectral density $\rho(\lambda)$ are poorly
reproduced by the guessed function. Apparently this is due to the contribution to $\rho(\lambda)$
from complex tree-like graphs as well as from non-tree-like graphs, which are present in the
ensemble for given $\eps$ as one can see from the list of graphs in the \fig{fig:1}.

Below we formulate two conjectures. In the first we propose an equation for the position of
sequential (in heights) peaks, while in the second we give an expression for the shape of an
outbound enveloping curve for the spectral density. These conjectures are based on a combination of
rigorous results concerning the spectra of tree-like graphs obtained in
\cite{rojo1,rojo2,rojo3,sen}, with some observations of ultrametric properties of a Dedekind
modular $\eta$-function.

\noindent {\bf Conjecture 1}. Any eigenvalue, $\lambda_i$, contributing to the spectral density,
$\rho(\lambda)$ for $p=1/N+\eps$ ($N\gg 1$ and $\eps\to 0$), can be uniquely associated with the
rational number, $\disp \frac{p_i}{q_i}$, where $p_i$ and $q_i$ are coprimes, as follows
\be
\lambda_i = -2\sqrt{d}\cos\frac{\pi p_i}{q_i}
\label{eq:03}
\ee
where $d$ is the maximal vertex degree of trees. In diluted sparse regime one has mainly $d=1$
(linear chains) or $d=2$ (star-like graphs).

The outbound enveloping peaks, $\lambda_1,\lambda_2,...$ (see \fig{fig:2}) are located at
\be
\lambda_m=2\cos\frac{\pi}{m+1} \qquad (m=1,2,...)
\label{eq:03a}
\ee

\noindent {\bf Conjecture 2}. The function $f(\lambda) = \rho^{1/\nu}(\lambda)$ for $\nu=4$
brings any subsequence of \emph{monotonically} increasing (or monotonically decreasing) peaks into
a linear form $f(\lambda) = a + b \lambda$, where $a$ and $b$ are subsequence-dependent constants.
In particular, under such a transformation, the spectral density (and the outbound enveloping shape
as well) acquires the triangular shape (see the \fig{fig:5}). Thus, the outbound enveloping curve
for the spectral density, $\rho(\lambda)$, has the following parametric representation ($0\le
\lambda\le 2$):
\be
\rho(\lambda)=c\, (2-\lambda)^{\nu}=\left(2-2\cos\frac{\pi}{m+1}\right)^{\nu} \qquad (m=1,2,...)
\label{eq:04}
\ee
where $c$ is some constant. The comments concerning this conjecture are given in the Section
\ref{back}.

\section{Hints beyond the conjectures}

The positions of peaks in the spectrum can be found by using the results of the works
\cite{rojo2,rojo3}, where the spectral properties of trees have been investigated. In particular,
it has been found in \cite{rojo3} that the spectrum of a regular tree-like graph is defined by the
eigenvalues of the three-diagonal symmetric matrix. The Conjecture 1 is based on the supposition
that the set of outbound peaks in the eigenvalue density $\rho(\lambda)$ is the set of maximal
eigenvalues $\{\lambda_{\rm max}^{(1)},\,\lambda_{\rm max}^{(2)},...\}$, where $\disp \lambda_{\rm
max}^{(m)}=2\cos\frac{\pi}{m+1}$ is the maximal eigenvalue of an $m$-vertex linear subchain.
Examining the \fig{fig:2} one can see that the spectral density $\rho(\lambda)$ extends beyond the
value $\lambda=2$ which is the terminal eigenvalue for linear chains. This means that near the
tails of the distribution the tree-like graphs and graphs with loops become dominant -- see the
Section \ref{back} for more discussion. Apparently these subgraphs cannot be eliminated by
decreasing $\eps$.

The Conjecture 2 is more involved and is motivated by some similarities between the spectral
density $\rho(\lambda)$ and the ultrametric structure of "continuous trees" isometrically embedded
in hyperbolic domains \cite{nech1}. Below we summarize some facts concerning the "isometric
continuous trees".

\subsection{Ultrametric structure of isometric Cayley trees}

\begin{figure}[ht]
\epsfig{file=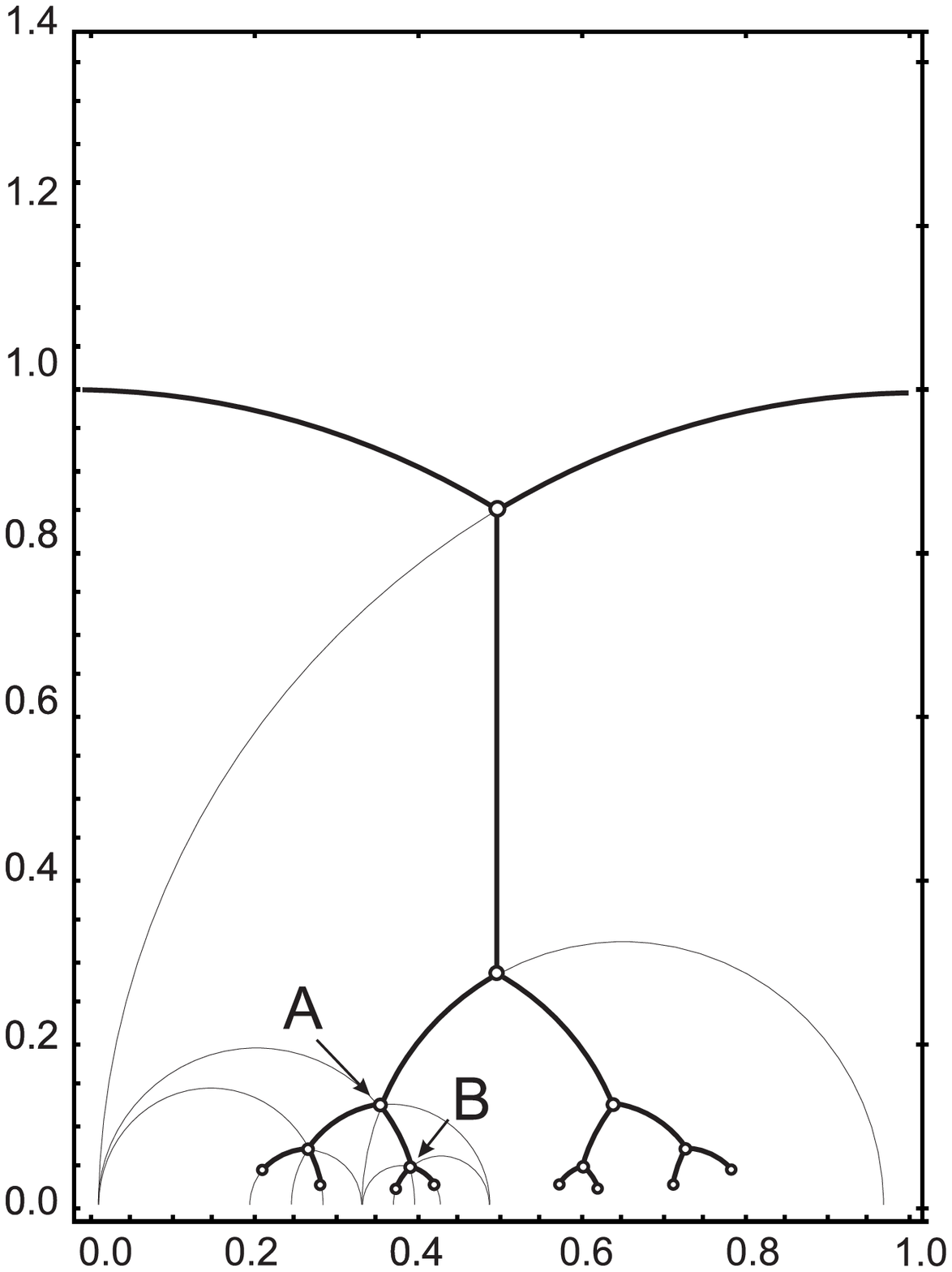,width=4cm} \quad \epsfig{file=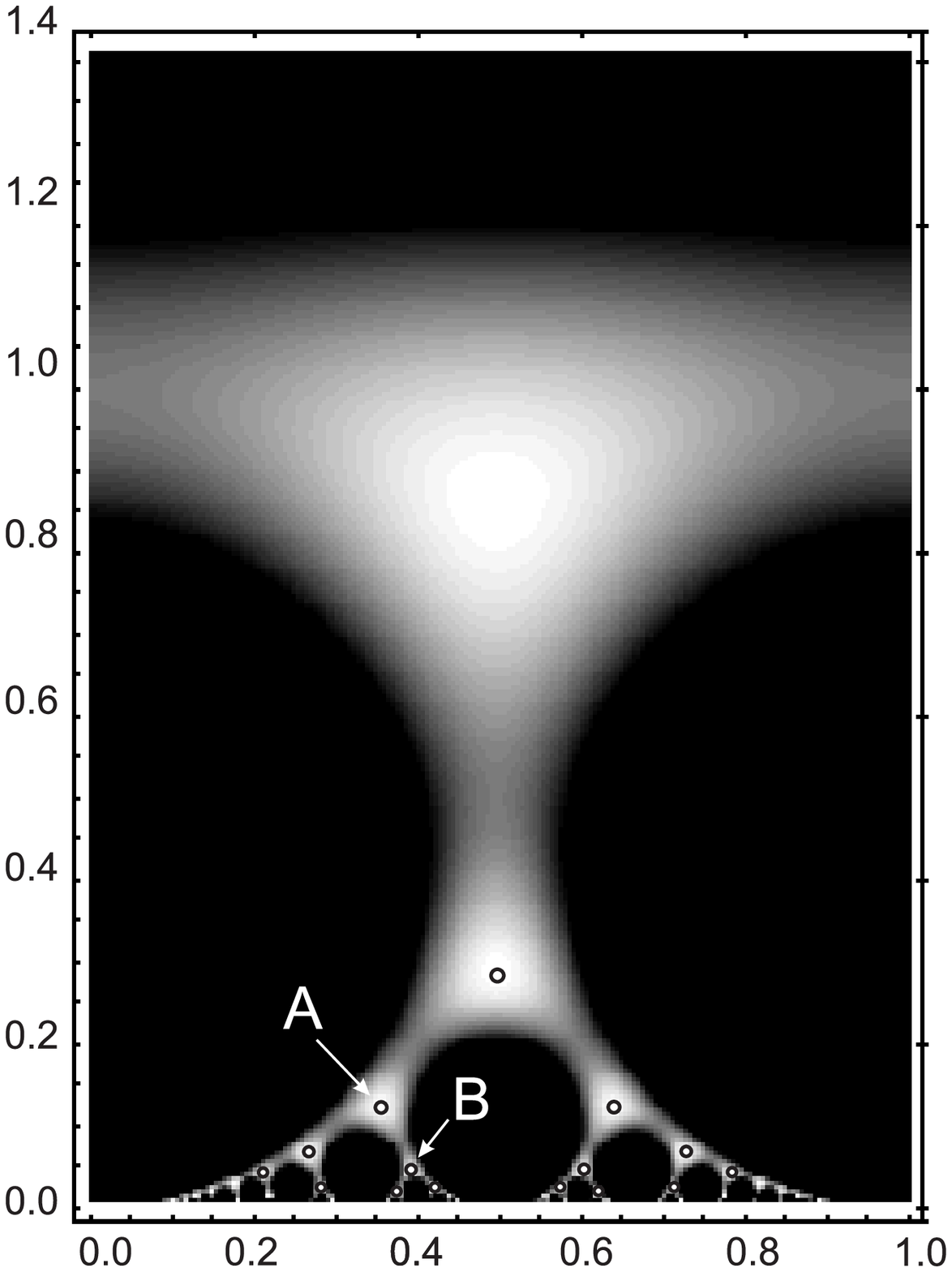,width=4cm} \quad
\epsfig{file=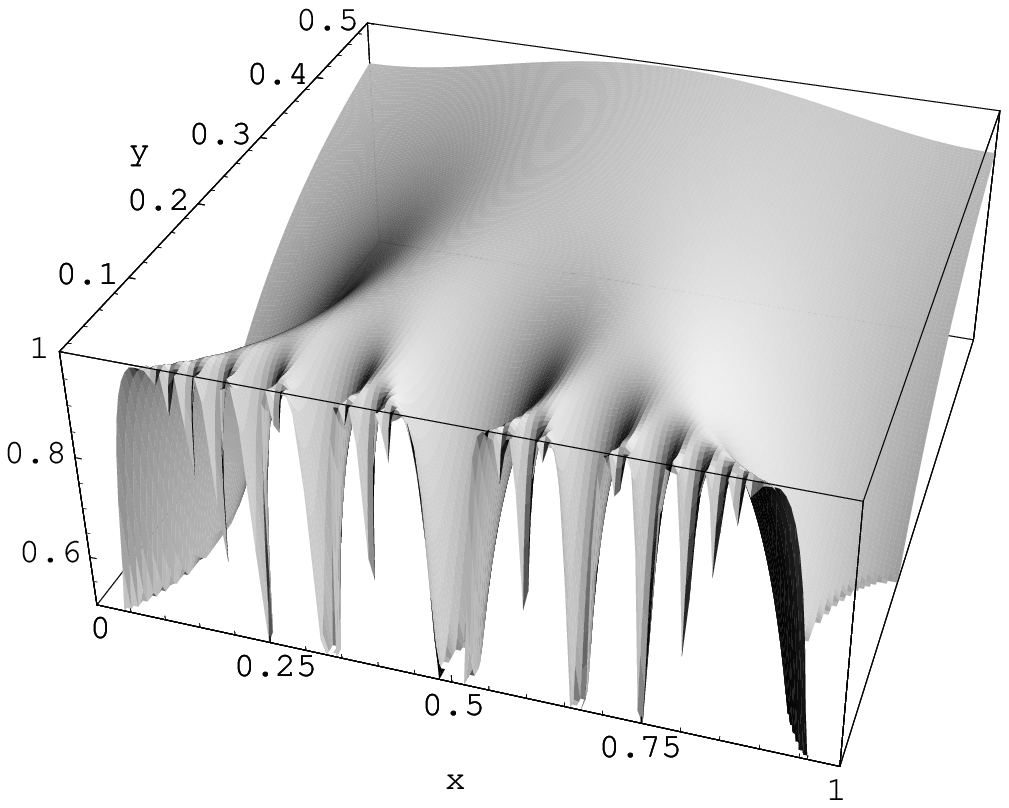,width=6cm}
\caption{Left: 3-branching Cayley tree isometrically embedded in Poincar\'e hyperbolic upper
half-plane ${\cal H}$. Center: Density plot of the function $\tilde{f}(z)$ (see the text) in the
rectangle $\{0\le {\rm Re}\,z\le 1,\; 0.01\le {\rm Im}\,z\le 1.4\}$. Right: Relief of the function
$\tilde{f}(z)$ in the rectangle $\{0\le {\rm Re}\,z\le 1,\; 0.04\le {\rm Im}\,z\le 1.4\}$.}
\label{fig:3}
\end{figure}

Any regular Cayley tree, as an exponentially growing structure, cannot be isometrically embedded in
an Euclidean plane. The embedding of a Cayley tree ${\cal C}$ into the metric space is called
"isometric" if ${\cal C}$ covers that space, preserving all angles and distances. The Cayley tree
${\cal C}$ isometrically covers the surface of constant negative curvature (the Lobachevsky plane)
${\cal H}$. One of possible representations of ${\cal H}$, known as a Poincar\'e model, is the
upper half-plane ${\rm Im}\,z>0$ of the complex plane $z=x+iy$ endowed with the metric
$ds^2=\frac{dx^2+dy^2} {y^2}$ of constant negative curvature. In \cite{nech1} we have constructed
the "continuous" analog of the standard 3-branching Cayley tree by means of special (modular)
functions and have analyzed the structure of the barriers separating the neighboring valleys. In
particular, we have shown that due to specific properties of modular functions these barriers are
ultrametrically organized. The main ingredient of our construction was the function $f(z)$ defined
as follows:
\be
\tilde{f}(z)=C^{-1}\,|\eta(z)|\, ({\rm Im}\,z)^{1/4}
\ee
where $\eta(z)$ is the Dedekind $\eta$-function (see, for instance \cite{chand})
\be
\eta(z)=e^{\pi i z/12}\prod_{k=0}^{\infty}(1-e^{2\pi i k z}); \qquad {\rm Im}\,z>0
\label{eq:2}
\ee
The normalization constant $C=\left|\eta\left(\frac{1}{2}+i\frac{\sqrt{3}}{2}\right)\right|\,
\left(\frac{\sqrt{3}}{2}\right)^{1/4}=0.77230184... $ is chosen to fix the maximal value of the
function $\tilde{f}(z)$ equal to 1: $0<\tilde{f}(z)\le 1$ for any $z$ in the upper half-plane ${\rm
Im}\,z>0$. The function $\tilde{f}(z)$ has the following property: all the solutions of the
equation $\tilde{f}(z)-1=0$ define all the coordinates of the 3-branching Cayley tree isometrically
embedded into the upper half-plane ${\cal H}(z|{\rm Im}\,z>0)$. The corresponding Cayley tree and
the density plot of the function $\tilde{f}(z)$ in the region $\{0\le {\rm Re}\,z \le 1,\; 0.04\le
{\rm Im}\,z\le 1.4\}$ is shown in fig.\ref{fig:3}. It is noteworthy that the function
$Z(z)=\left[C\,\tilde{f}(z)\right]^{-2}$ is invariant with respect to the action of the modular
group $PSL(2,\mathbb{Z})$, namely, $Z(z) = Z(z+1)$ and $Z(z) = Z\left(-\frac{1}{z}\right)$. The 3D
relief of the function $f(z)$ is shown in figure \ref{fig:3}(right).

\begin{figure}[ht]
\epsfig{file=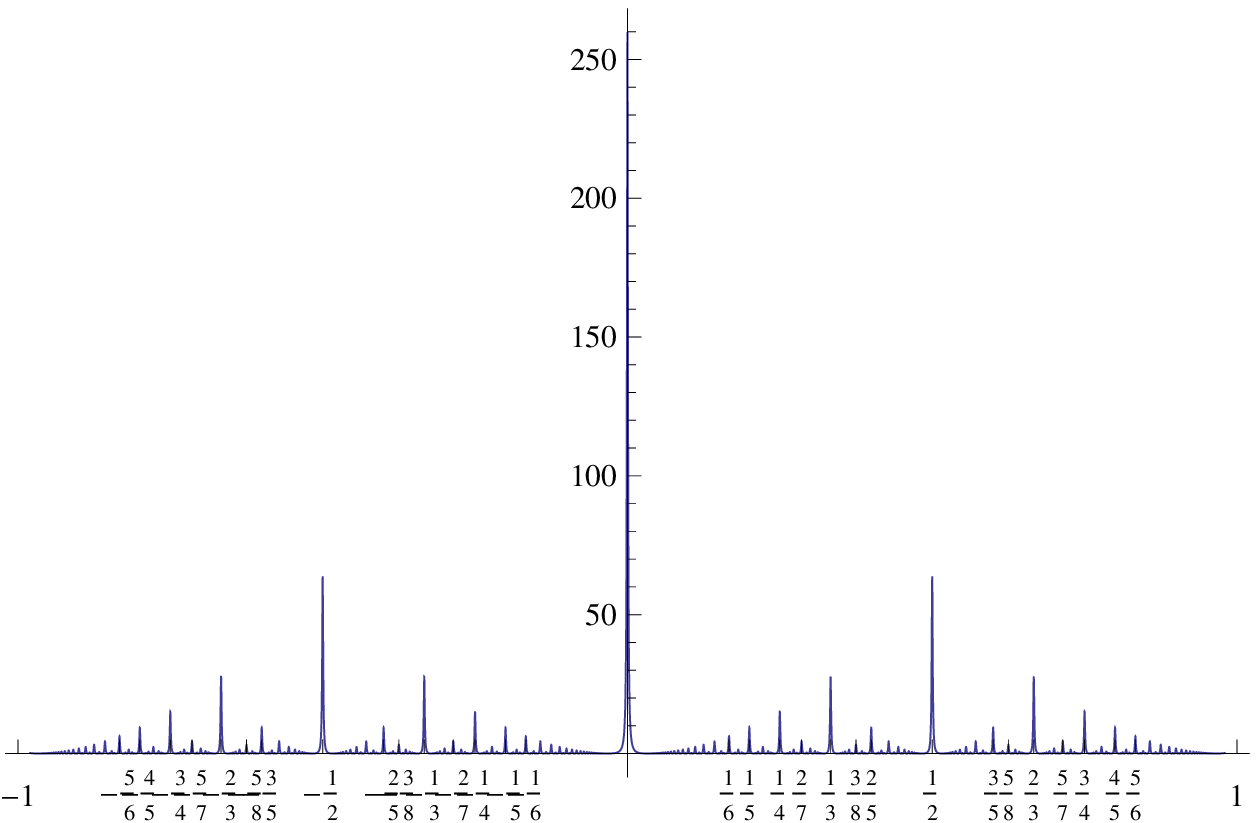,width=8cm} \epsfig{file=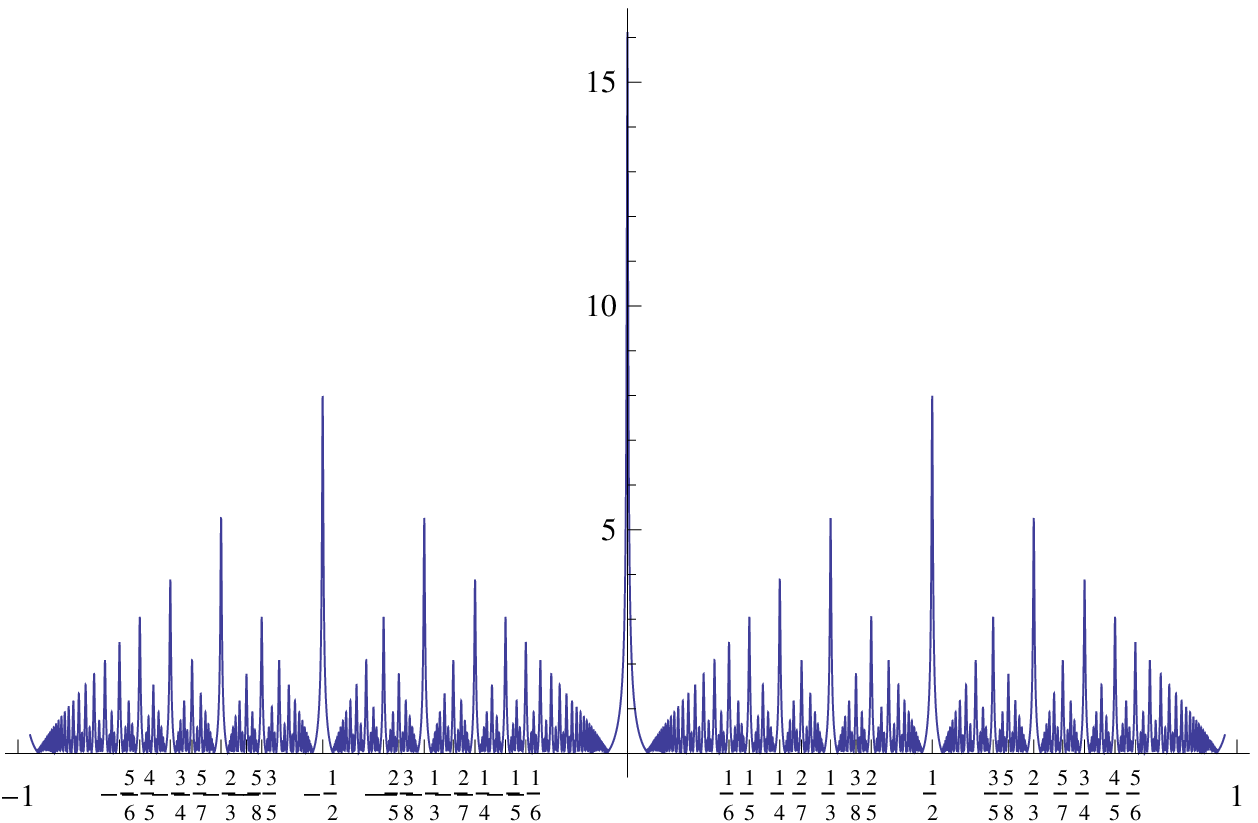,width=8cm}
\caption{Typical shape of the functions $v(x)=v(x|y=0.001)$ (left) and $\sqrt{v(x)}=
\sqrt{v(x|y=0.001)}$ (right).}
\label{fig:4}
\end{figure}

The "continuous tree-like structure" of hills separated by the valleys is clearly seen in the
\fig{fig:3}. Consider now the function
\be
v(x|y)=-\ln \tilde{f}(x|y); \qquad x={\rm Re}\,z;\; y={\rm Im}\,z
\label{eq:v}
\ee
The typical shape of the function $v(x|y)$, shown in the figure \fig{fig:4}(left), demonstrates the
ultrametric organization of the barriers separating the valleys. In the figure \fig{fig:4}(right)
we have drawn the relief of the function $u(x) = v^{1/2}(x)$. Such a transformation highlights the
"linearity" of growth of sequentially increasing and decreasing barriers. This fact can be exactly
proved using the algebraic properties of the Dedekind $\eta$-function (see the Appendix 1).

The function $v(x)=-\ln \tilde{f}(x)$, defined on the interval $0<x<1$, has the properties borrowed
from the structure of the underlying modular group $PSL(2,\mathbb{Z})$ acting in the half--plane
${\cal H}(z|y>0)$ \cite{magnus,beardon}. In particular: (i) the local maxima of the function $v(x)$
are located at the rational points; (ii) the highest barrier on a given interval $\Delta x =
[x_1,x_2]$ is located at a rational point $\frac{p}{q}$ with the lowest denominator $q$. On a given
interval $\Delta x = [x_1,x_2]$ there is only one such point. The locations of the barriers with
the consecutive heights on the interval $\Delta x$ are organized according to the group operation:
\be
\frac{p_1}{q_1}\oplus \frac{p_2}{q_2} = \frac{p_1+p_2}{q_1+q_2}
\label{eq:rule}
\ee
The \fig{fig:4} clarify this statement. The highest barriers on the interval $0\le x\le 1$ are
located at the points $x_0=0$ and $x_1=1$. Rewriting $0$ and $1$ correspondingly as $\frac{0}{1}$
and $\frac{1}{1}$ we can find the point $x_2$ of location of the barrier with the next maximal
hight. Namely, $x_2=\frac{0}{1}\oplus \frac{1}{1}=\frac{0+1}{1+1}=\frac{1}{2}$. Continuing this
construction we arrive at the hierarchical  structure of barriers located at rational points
organized in the Farey sequence.

\subsection{Back to the spectral density of diluted sparse matrices}
\label{back}

Now we are in position to give a hint why the function $g(\lambda)$ has appeared in
\eq{eq:comp1}--\eq{eq:comp2}. The argument, $\frac{1}{\pi} \arccos \frac{\lambda}{2}$, of the
function $f(...)$ in \eq{eq:comp1} is just the inverted expression of \eq{eq:03}. This expression
projects back the eigenvalues $\lambda_i$ to the rational numbers $\disp \frac{p_i}{q_i}$.

\begin{figure}[ht]
\epsfig{file=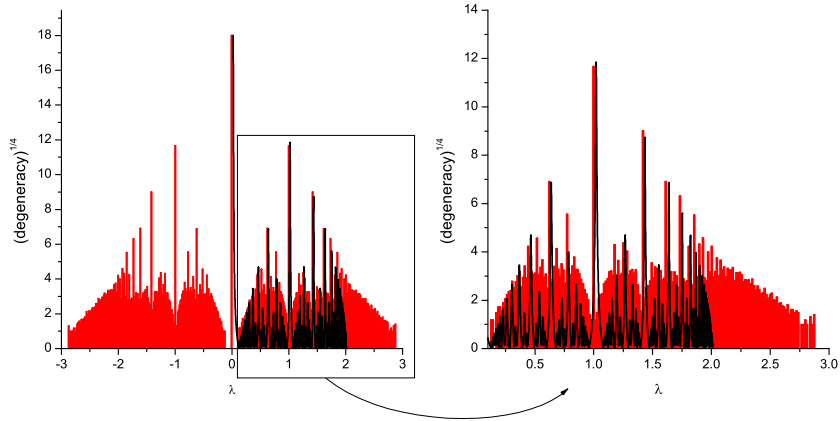,width=11cm}
\caption{The same distributions as in \fig{fig:2-1}, but the plotted functions are:
$\rho^{1/4}(\lambda)$ (red) and $g^{1/4}(\lambda)$ (black).}
\label{fig:5}
\end{figure}

We have noted that the spectral density $\rho(\lambda)$ looks pretty much similar as the function
$v^2(\lambda|y)$ taken at some fixed $0<y\ll 1$. Just this comparison is plotted in the
\fig{fig:2-1}. The appearance of the exponent $\nu=4$ in \eq{eq:04} is the consequence of the guess
\eq{eq:comp1}. Namely, since $g(\lambda) \sim v^2(\lambda)$ and the function $v^{1/2}(\lambda)$
brings the Dedekind relief to the linear shape, we conclude that the function $g^{1/4}(\lambda)$
makes the shape of the transformed spectral density $\rho^{1/4}(\lambda)$ also linear -- see the
\fig{fig:5}. The deviation from the linearity at the tails of the Dedekind analytic disappear in
the limit $y\to 0$ as one can see from the \eq{ap:09}.

The discrepancy between tails of the functions $\rho(\lambda)$ and $g(\lambda)$ are due to
essential contributions from large tree-like graphs, as well as from non-tree-like graphs in the
sparse matrix ensemble. According to \cite{rojo3}, the largest eigenvalue of the tree-like graph
can be roughly estimated \footnote{In \cite{rojo3} more refined estimate is given.} as
$\lambda_{\rm max}\le 2\sqrt{d}$ where $d$ is the maximal vertex degree of a tree. In the bulk of
the spectrum the contribution from the linear chain-like graphs (i.e. graphs with $d=2$) dominate,
while at the edge of the spectrum the contribution from tree-like graphs with $d>2$ become
essential.

Note however that heavy tails of the function $\rho^{1/4}(\lambda)$ beyond the terminal value
$|\lambda|=2$ for chain-like graphs, also demonstrate linear behavior (however with a slope
different from the slop in a bulk).

\section{Discussion}

In this letter we have conjectured that the spectral density, $\rho(\lambda)$ of an ensemble of
"diluted" random symmetric sparse Bernoulli matrices demonstrates in the bulk the structure which
can be roughly reproduced using the construction which involves the Dedekind modular
$\eta$-function.

Yet we have not any proof of the connection between $\rho(\lambda)$ and the Dedekind
$\eta$-function, however this similarity seems not occasional. Rephrasing the question of M. Kac
\cite{kac} as "Can one hear the shape of a tree?", we expect that some information about the
spatial structure of a tree is encoded in its spectrum. The Weyl's conjecture (which allows one to
estimate the size and the area of the surface by the total number of eigenvalues) applied to trees,
does not help much because the number of eigenvalues trivially coincides with the size of the
adjacency matrix of a tree, and the area of a tree is indistinguishable from the volume (both are
the number of a tree vertices). Nevertheless one sees that non-isomorphic trees have different sets
of eigenvalues. What is the spectral density, $\rho(\lambda,k)$ of ensemble of all trees of a given
number of vertices, $N$, and a fixed maximal backbone length, $k$? Despite some important results
are obtained in \cite{rojo3} for particular tree-like graphs, the full answer to this question is
still not known.

On the other side, the Dedekind function has appeared in some counting problems in hyperbolic
domains related to so-called "arithmetic chaos" considered by M. Gutzwiller and B. Mandelbrot
\cite{gut}. In particular they found the connection of the arithmetic function $\beta(\xi)$ which
maps some number $\xi\in[0,1]$, written as a continued fraction expansion
\be
\frac{1}{n_1+\disp\frac{1}{n_2+\ldots}}
\label{eq:cont}
\ee
to the real number $\beta$, whose binary expansion is made by the sequence of $n_1-1$ times 0,
followed by $n_2$ times 1, then $n_3$ times 0, and so on. In the work \cite{series} C. Series
pointed out the that the continued fraction expansion \eq{eq:cont} is related to the following
counting problem in the hyperbolic upper half-plane ${\cal H}(z|{\rm Im}\,z>0)$ -- see \fig{fig:3}.
Take the root point of the Cayley tree isometrically embedded in ${\cal H}$ and compute how many
vertices (images) of the root point lie in the interval $[0,\xi]$. Therefore $\beta(\xi)$ is
proportional to a fractal "invariant measure" (i.e. to the number of vertices lying in the interval
$[0,\xi]$). The counting of vertices, whose real part lie in $[0,\xi]$, is equivalent to counting
the number of maxima of the function $|\eta(x+i0^+)|$ in the same interval.

To summarize, let us emphasize that apparently the question considered in this letter lies at the
edge of the spectral theory of tree-like graphs and the "arithmetic chaos" dealing with isometric
embedding of these graphs into the hyperbolic domains. It seems that the near future will show if
our guess is correct or not.

I would like to thank Eugene Bogomolny for providing me references on spectral structure of
tree-like graphs and to Olga Valba for useful discussions.

\begin{appendix}

\section{"Shape linearity" of the relief $u(x)=\sqrt{-\ln f(x)}$.}

Let the function $f(z)$ for ${\rm Im}\, z>0$ be:
\be
f(z) = \left|\eta(z) \right| \left({\rm Im}\, z\right)^{1/4}
\label{ap:01}
\ee
where
\be
\eta(z)=e^{\pi i z/12}\prod_{k=1}^{\infty}(1-e^{2\pi i k z}); \quad {\rm Im}\,z>0
\label{ap:02}
\ee
is the Dedekind $\eta$--function.

The following duality relation is valid for $f(z)$:
\be
f\left(\frac{p}{q} + iy\right) = f\left(\frac{s}{q}+\frac{i}{q^2y} \right)
\label{ap:03}
\ee
if and only if $p$ and $q$ are coprime, $s=(1+q r)/p$ and $r$ are integers.

Using \eq{ap:03} we can obtain the asymptotics of $\eta(z)$ when ${\rm Im}\,z\to 0^+$. Take into
account the relation of Dedekind $\eta$--function with Jacobi elliptic functions:
\be
\vartheta_1'(0,e^{\pi i z})=\eta^3(z)
\label{ap:04}
\ee
where
\be
\vartheta_1'(0,e^{\pi i z})\equiv \frac{d\vartheta_1(u,e^{\pi i z})}{du}\bigg|_{u=0}= e^{\pi i
z/4}\sum_{n=0}^{\infty}(-1)^n (2n+1)e^{\pi i n(n+1)z}
\label{ap:05}
\ee
Rewrite \eq{ap:03} as follows:
\be
\left|\eta\left(\frac{p}{q} + iy\right)\right|=\frac{1}{(q y)^{1/2}}\left|\eta\left(\frac{s}{q} +
\frac{i}{q^2y}\right)\right|
\label{ap:06}
\ee
Applying \eq{ap:04}--\eq{ap:05} to the r.h.s of \eq{ap:06}, we get:
\be
\left|\eta\left(\frac{p}{q} + iy\right)\right|=\frac{1}{(q y)^{1/2}}\left|e^{\frac{\pi}{4} i
\left(\frac{s}{q}+\frac{i}{q^2y}\right)}\sum_{n=0}^{\infty}(-1)^n (2n+1)e^{\pi i
n(n+1)\left(\frac{s}{q}+\frac{i}{q^2y} \right)} \right|^{1/3}
\label{ap:07}
\ee
Eq.\eq{ap:07} enables us to extract the asymptotics of $\eta(z)$ at ${\rm Im}\,z\to 0^+$:
\be
\left|\eta\left(\frac{p}{q} + iy\right)\right|\Bigg|_{y\to 0^+}=\frac{1}{(q y)^{1/2}} e^{-\pi/(12
q^2y)}
\label{ap:08}
\ee
Thus,
\be
\sqrt{-\ln \left|\eta\left(\frac{p}{q} + iy\right)\right|}\Bigg|_{y\to 0^+} =
\frac{1}{q}\sqrt{\frac{\pi}{12y}} + \frac{1}{4}\left(\ln q + \ln y\right)
\label{ap:09}
\ee
Denoting $\frac{1}{q}\equiv x$, we see from \eq{ap:09} the dominant contribution from the linear
term in $x$ as $y\to 0^+$ on the "coprime subsequences".

\end{appendix}

\end{document}